\documentclass[reprint,noshowpacs,noshowkeys,prd,balancelastpage,nofootinbib]{revtex4}
\usepackage{amsfonts}
\usepackage{mdframed}
\usepackage{amssymb}
\usepackage{footnote}
\usepackage{amsmath}
\usepackage{graphicx}
\usepackage{float}
\usepackage[font={footnotesize,it}]{caption}
\usepackage[utf8]{inputenc}
\usepackage{natbib}
\usepackage{tikz}
\usepackage{tikz-3dplot}
\usepackage[colorlinks=true,
            linkcolor=purple,
            urlcolor=purple,
            citecolor=blue]{hyperref}

\begin{document}

\title{Axially Symmetric Exponential Metric }
\author{S. Habib Mazharimousavi}
\email{habib.mazhari@emu.edu.tr}
\affiliation{Department of Physics, Faculty of Arts and Sciences, Eastern Mediterranean
University, Famagusta, North Cyprus via Mersin 10, T\"{u}rkiye}
\date{\today }

\begin{abstract}
We revisit the Yilmaz Exponential Metric (YEM), which was recently shown to
describe a traversable wormhole, and construct an axially symmetric
generalization strictly within the framework of general relativity. Starting
from a deformed spherically symmetric ansatz, we introduce a two-parameter
family of Axially Symmetric Exponential Metric (ASEM) solutions that
smoothly interpolate between the YEM and the Curzon-Chazy spacetime geometry.
\end{abstract}

\maketitle

\section{Introduction}

Recently, in a detailed study, Boonserm, Ngampitipan, Simpson, and Visser
demonstrated that the long-standing Yilmaz Exponential Metric (YEM)
represents a traversable wormhole solution \cite{E1}. To the best of our
knowledge, the YEM first appeared in a 1954 publication by Papapetrou (in
German) \cite{P1}, with the line element 
\begin{equation}
ds^{2}=-e^{\left( -\frac{2M}{r}\right) }dt^{2}+e^{\left( +\frac{2M}{r}%
\right) }\left[ dr^{2}+r^{2}d\theta ^{2}+r^{2}\sin ^{2}\theta d\phi ^{2}%
\right] .  \label{1.1}
\end{equation}%
This one-parameter solution to Einstein's field equations is singular and
non-black hole in nature. It is supported by a matter field with an
energy-momentum tensor given by 
\begin{equation}
T_{\mu }^{\nu }=\frac{M^{2}}{r^{4}e^{\frac{2M}{r}}}diag\left[ 1,-1,1,1\right]
.  \label{1.2}
\end{equation}%
As Papapetrou noted in the abstract of \cite{P1}, his aim was to improve
upon two earlier works published under the title A New Theory of the
Gravitational Field: I \cite{P2} and II \cite{P3}. The metric in (\ref{1.1})
quickly attracted attention, particularly from Yilmaz, who in 1958 proposed,
under a provocative title and abstract for its time, that the exponential
metric describes a generally covariant scalar field theory of gravity \cite%
{Y1}.

In 1971, Yilmaz further advanced his New Theory of Gravitation \cite{Y2}, in
which the metric tensor was no longer the gravitational potential itself but
instead a functional of a spin-2 field. This theory was subsequently
developed further in \cite{Y3}. Following these efforts, the solution became
known as the Yilmaz Exponential Metric (YEM) \cite{E2}.

It is important to distinguish between the Yilmaz theory of gravitation and
the YEM. The latter is merely one solution that arises within the framework
of the former. The general form of the solution in Yilmaz theory is given by%
\begin{equation}
ds^{2}=-e^{-2\phi }dt^{2}+e^{2\phi }\left[ dr^{2}+r^{2}d\theta
^{2}+r^{2}\sin ^{2}\theta d\phi ^{2}\right] ,  \label{R0}
\end{equation}%
where the scalar field $\phi $\ satisfies the corresponding field equations
in the theory. For a single point mass, one obtains $\phi =\frac{M}{r}$\ 
\cite{Y1}. In the weak-field limit, Yilmaz's classical field theory agrees
with general relativity. Indeed, in the limit $\frac{M}{r}\rightarrow 0,$
the metric (\ref{1.1}) reduces to%
\begin{equation}
ds_{YEM}^{2}\rightarrow -\left( 1-\frac{2M}{r}\right) dt^{2}+\left( 1+\frac{%
2M}{r}\right) \left\{ dr^{2}+r^{2}d\theta ^{2}+r^{2}\sin ^{2}\theta d\phi
^{2}\right\} ,\text{ as }\frac{M}{r}\rightarrow 0  \label{I1}
\end{equation}%
which matches the weak-field limit of the Schwarzschild black hole expressed
in isotropic coordinates%
\begin{equation}
ds_{Sch}^{2}=-\left( \frac{1-\frac{M}{2r}}{1+\frac{M}{2r}}\right)
^{2}dt^{2}+\left( 1+\frac{M}{2r}\right) ^{4}\left\{ dr^{2}+r^{2}d\theta
^{2}+r^{2}\sin ^{2}\theta d\phi ^{2}\right\} .  \label{I2}
\end{equation}%
Yilmaz's theory and the YEM have been the subject of significant attention
in both early works \cite{E3,E4,E5,E6,E7,E8,E9,E10,E11,E12,E13,E14} and more
recent studies \cite{E15,E16,E17,E18,E19,E20,E21,E22,E23,E24,E25}, receiving
both support and criticism. One of the most recent and notable
investigations is \cite{E1}, which treats the exponential metric as a fixed
background geometry, independent of the Yilmaz theory. The authors
rigorously analyze the conditions for traversable wormholes and conclude
that the spacetime described by the YEM is an asymptotically flat,
traversable wormhole \cite{T1,T2,T3,T4,T5}.

According to \cite{E1}, this wormhole is asymmetric with respect to its
throat, which occurs at $r=M,$ where the area of the spatial hypersurface
reaches a minimum. All metric components remain finite and nonzero at the
throat, confirming the flare-out condition is satisfied. However, to
maintain traversability, the wormhole must be supported by exotic matter. As
shown in (\ref{1.2}), the energy density is negative: 
\begin{equation}
\rho =-\frac{M^{2}}{r^{4}}e^{-\frac{2M}{r}}<0,  \label{N1}
\end{equation}
implying that, except for the null energy condition, all classical energy
conditions (weak, strong) are violated. Thus, the exponential metric
wormhole is powered by exotic matter.

Given the extensive literature already devoted to the YEM, we do not aim to
expand upon its properties in this Letter. Instead, our focus shifts toward
an axially symmetric generalization of the YEM, strictly within the
framework of general relativity.

To this end, we revisit the Curzon-Chazy spacetime (CCS) \cite{Cur,Cha},
originally expressed in cylindrical coordinates as%
\begin{equation}
ds^{2}=-e^{2\psi }dt^{2}+e^{-2\psi }\left( e^{\gamma }\left( d\rho
^{2}+dz^{2}\right) +\rho ^{2}d\phi ^{2}\right) ,  \label{I3}
\end{equation}%
with the metric functions%
\begin{equation}
\psi =-\frac{M}{\sqrt{\rho ^{2}+z^{2}}},\text{ \ \ }\gamma =-\frac{M^{2}\rho
^{2}}{\rho ^{2}+z^{2}}.  \label{I4}
\end{equation}%
We reformulate the Einstein field equations in spherical coordinates and
solve them exactly to investigate possible extensions of this vacuum
solution. Our ultimate objective is to derive a metric that satisfies
Einstein's equations and encompasses both the Yilmaz Exponential Metric and
the Curzon-Chazy spacetime as limiting cases.

\section{Revisiting the Curzon-Chazy Spacetime in Spherical Coordinates}

Inspired by the Yilmaz Exponential Metric (YEM) and the Curzon--Chazy
spacetime (CCS), we propose a deformed line element of the form%
\begin{equation}
ds^{2}=-\psi \left( r\right) dt^{2}+\frac{1}{\psi \left( r\right) }\left\{
\xi \left( r,\theta \right) \left( dr^{2}+r^{2}d\theta ^{2}\right)
+r^{2}\sin ^{2}\theta d\phi ^{2}\right\} ,  \label{2.1}
\end{equation}%
where $\psi \left( r\right) $ and $\xi \left( r,\theta \right) $ are metric
functions to be determined by solving Einstein's vacuum field equations. A
comparison between expressions (\ref{2.1}) and (\ref{1.1}) reveals that the
deformation arises through the function $\xi \left( r,\theta \right) ,$
which modifies the radial and polar length elements.

The vacuum Einstein field equations can be written in terms of the Ricci
tensor as:%
\begin{equation}
R_{\mu \nu }=0.  \label{2.2}
\end{equation}%
From the metric (\ref{2.1}), the non-zero components of the Ricci tensor are
computed as%
\begin{equation}
R_{tt}=\frac{-\psi ^{\prime 2}r+\psi \psi ^{\prime \prime }r+2\psi \psi
^{\prime }}{2\xi r},  \label{2.3}
\end{equation}%
\begin{equation}
R_{rr}=-\frac{\xi _{,\theta }}{2\xi r^{2}\tan \theta }+\frac{\psi ^{\prime
\prime }}{2\psi }-\frac{\psi ^{\prime 2}}{\psi ^{2}}+\frac{\psi ^{\prime }}{%
r\psi }+\frac{\xi _{,r}^{2}}{2\xi ^{2}}-\frac{\xi _{,rr}}{2\xi }+\frac{\xi
_{,\theta }^{2}}{2r^{2}\xi ^{2}}-\frac{\xi _{,\theta \theta }}{2r^{2}\xi },
\label{2.4}
\end{equation}%
\begin{equation}
R_{r\theta }=\frac{\xi _{,\theta }}{2r\xi }+\frac{\xi _{,r}}{2\xi \tan
\theta },  \label{2.5}
\end{equation}%
\begin{equation}
R_{\theta \theta }=\frac{\xi _{,\theta }}{2\xi \tan \theta }-\frac{r^{2}\psi
^{\prime 2}}{2\psi ^{2}}+\frac{r^{2}\psi ^{\prime \prime }}{2\psi }+\frac{%
r\psi ^{\prime }}{\psi }-\frac{r\xi _{,r}}{\xi }-\frac{r^{2}\xi _{,rr}}{2\xi 
}+\frac{r^{2}\xi _{,r}^{2}}{2\xi ^{2}}+\frac{\xi _{,\theta }^{2}}{2\xi ^{2}}-%
\frac{\xi _{,\theta \theta }}{2\xi },  \label{2.6}
\end{equation}%
and%
\begin{equation}
R_{\phi \phi }=\frac{r\sin ^{2}\theta \left( -\psi ^{\prime 2}r+\psi \psi
^{\prime \prime }r+2\psi \psi ^{\prime }\right) }{2\xi \psi ^{2}}.
\label{2.7}
\end{equation}%
Here, a prime denotes differentiation with respect to $r$, while partial
derivatives are expressed as $Q_{,x}=\frac{\partial Q}{\partial x}$.
Starting with the first field equation, $R_{tt}=0,$ we find the exact
solution for $\psi \left( r\right) $ as 
\begin{equation}
\psi \left( r\right) =C_{2}e^{\frac{C_{1}}{r}},  \label{2.8}
\end{equation}%
where $C_{1}$ and $C_{2}$ are integration constants. The off-diagonal
component $R_{r\theta }=0$ leads to a class of solutions for the deformation
function $\xi \left( r,\theta \right) ,$ given by%
\begin{equation}
\xi \left( r,\theta \right) =\xi \left( \frac{\sin \theta }{r}\right) ,
\label{2.9}
\end{equation}%
where $\xi \left( z\right) $ is an arbitrary, well-defined function of $z=%
\frac{\sin \theta }{r}.$ Substituting equations (\ref{2.8}) and (\ref{2.9})
into the next field equation $R_{rr}=0,$ we find an explicit solution for $%
\xi \left( z\right) $ 
\begin{equation}
\xi \left( z\right) =C_{3}z^{C_{4}}e^{-\frac{C_{1}^{2}}{4}z^{2}},
\label{2.10}
\end{equation}%
which in terms of $r$ and $\theta $ becomes%
\begin{equation}
\xi \left( r,\theta \right) =C_{3}\left( \frac{\sin \theta }{r}\right)
^{C_{4}}e^{-\left( \frac{C_{1}\sin \theta }{2r}\right) ^{2}}.  \label{2.11}
\end{equation}%
Finally, applying the equation $R_{\theta \theta }=0$ determines that $%
C_{4}=0,$ yielding the final line element%
\begin{equation}
ds^{2}=-C_{2}e^{\frac{C_{1}}{r}}dt^{2}+\frac{1}{C_{2}e^{\frac{C_{1}}{r}}}%
\left\{ C_{3}e^{-\left( \frac{C_{1}\sin \theta }{2r}\right) ^{2}}\left(
dr^{2}+r^{2}d\theta ^{2}\right) +r^{2}\sin ^{2}\theta d\phi ^{2}\right\} .
\label{2.12}
\end{equation}%
This solution contains three integration constants. Through a suitable
rescaling 
\begin{equation}
r\rightarrow \sqrt{\frac{C_{2}}{C_{3}}}r,t\rightarrow \frac{1}{\sqrt{C_{2}}}%
t,\phi \rightarrow \sqrt{C_{3}}\phi ,\text{ and }C_{1}=2M\sqrt{\frac{C_{2}}{%
C_{3}}}
\end{equation}%
the metric (\ref{2.12}) simplifies to a one-parameter solution%
\begin{equation}
ds^{2}=-e^{\left( -\frac{2M}{r}\right) }dt^{2}+e^{\left( +\frac{2M}{r}%
\right) }\left[ e^{\left( -\frac{M^{2}}{r^{2}}\sin ^{2}\theta \right)
}\left( dr^{2}+r^{2}d\theta ^{2}\right) +r^{2}\sin ^{2}\theta d\phi ^{2}%
\right] .  \label{2.13}
\end{equation}%
Here, $M$ is the only remaining parameter, which, as will be shown,
represents the mass of the solution. The metric (\ref{2.13}) is singular at $%
r=0$ and asymptotically flat as $r\rightarrow \infty .$ In the asymptotic
limit%
\begin{equation}
ds^{2}\rightarrow -\left( 1-\frac{2M}{r}\right) dt^{2}+\left( 1+\frac{2M}{r}%
\right) \left\{ dr^{2}+r^{2}d\theta ^{2}+r^{2}\sin ^{2}\theta d\phi
^{2}\right\} \text{ as }r\rightarrow \infty  \label{R1}
\end{equation}%
which matches the asymptotic behavior of the Schwarzschild metric in
isotropic coordinates, as given in (\ref{I2}). This confirms that $M$
corresponds to the total mass of the solution. The only singularity lies at $%
r=0,$ which is evident from the Kretschmann scalar 
\begin{equation}
\mathcal{K}=-\frac{16M^{2}e^{2\left( -\frac{2M}{r}+\frac{M^{2}}{r^{2}}\sin
^{2}\theta \right) }\left[ M^{2}\left( 3r^{2}+M^{2}-3Mr\right) \cos
^{2}\theta -M^{4}-6M^{2}r^{2}+6Mr^{3}+3M^{3}r-3r^{4}\right] }{r^{10}}.
\label{2.14}
\end{equation}%
In the absence of an event horizon, this singularity is naked. When $M=0,$
the solution reduces to flat Minkowski spacetime. Our detailed calculations
have shown that, in spherical coordinates with geometry described by the
line element (\ref{2.1}), the only vacuum solution is the CCS. In the
following section, we generalize this configuration to a non-vacuum geometry.

\section{A 2-parameter solution}

In this section, we address the following question: The line element given
in (\ref{2.1}) is a direct extension of the Yilmaz Exponential Metric (YEM)
in (\ref{1.1}), and it reduces to (\ref{1.1}) when $\xi \left( r,\theta
\right) \rightarrow 1$. How, then, can the YEM, associated with a
distributed source as described in (\ref{2.1}), have such a source, while
the CCS in (\ref{2.13}), which is a geometrically deformed case of (\ref{2.1}%
), represents a vacuum solution?

To resolve this apparent paradox, we must propose a more general extension
to the YEM. We introduce an additional exact solution belonging to the same
class as (\ref{2.1}), with the specific choice $\psi \left( r\right) =1$\
and $\xi \left( r,\theta \right) =e^{\left( -\frac{B^{2}}{r^{2}}\sin
^{2}\theta \right) }.$\ This yields the following line element and
corresponding energy-momentum tensor:%
\begin{equation}
ds^{2}=-dt^{2}+e^{\left( -\frac{B^{2}}{r^{2}}\sin ^{2}\theta \right) }\left(
dr^{2}+r^{2}d\theta ^{2}\right) +r^{2}\sin ^{2}\theta d\phi ^{2},  \label{R3}
\end{equation}%
and%
\begin{equation}
T_{\mu }^{\nu }=\frac{B^{2}e^{\frac{B^{2}}{r^{2}}\sin ^{2}\theta }}{r^{4}}%
diag\left[ -1,1,-1,-1\right] ,  \label{R4}
\end{equation}%
respectively. The metric (\ref{R3}) is asymptotically flat as $r\rightarrow
\infty $\ and exhibits a naked singularity at $r=0,$\ which is supported by
the energy-momentum tensor (\ref{R4}). This source satisfies the null, weak,
and strong energy conditions.

We now introduce an interpolating spacetime that connects the YEM in (\ref%
{1.1}) and the solution in (\ref{R3}), given by the two-parameter metric:%
\begin{equation}
ds^{2}=-e^{\left( -\frac{2M}{r}\right) }dt^{2}+e^{\left( +\frac{2M}{r}%
\right) }\left[ e^{\left( -\frac{B^{2}}{r^{2}}\sin ^{2}\theta \right)
}\left( dr^{2}+r^{2}d\theta ^{2}\right) +r^{2}\sin ^{2}\theta d\phi ^{2}%
\right] .  \label{R5}
\end{equation}%
The corresponding energy-momentum tensor for this spacetime is%
\begin{equation}
T_{\mu }^{\nu }=\frac{e^{\frac{B^{2}}{r^{2}}\sin ^{2}\theta }\left(
M^{2}-B^{2}\right) }{r^{4}e^{\frac{2M}{r}}}diag\left[ 1,-1,1,1\right] .
\label{R6}
\end{equation}%
This solution (\ref{R5}) represents a two-parameter extension of the YEM and
hereafter will be called Axially Symmetric Exponential Metric (ASEM). It
reduces to (\ref{1.1}) in the limit $B=0,$\ and to (\ref{R3}) when $M=0.$\
Moreover, depending on the sign of $M^{2}-B^{2}$, the source term (\ref{R6})
can describe either a physical or exotic anisotropic fluid. Notably, the CCS
given in (\ref{2.13}) corresponds to the particular case $M^{2}-B^{2}=0,$\
which is the unique configuration for which $R_{\mu }^{\nu }=0$\ and
consequently, $T_{\mu }^{\nu }=0.$

Therefore, the answer to the question posed at the beginning of this section
is that the direct and proper extension of the YEM in (\ref{1.1}) is the
two-parameter solution presented in (\ref{R5}), which is not sourceless. The
solution obtained by enforcing $R_{\mu }^{\nu }=0,$\ as in (\ref{2.13}), is
simply a special case of the more general extension in (\ref{R5}). In other
words, both the YEM (\ref{1.1}) and the CCS (\ref{2.13}) are specific
configurations within the broader class of spacetimes described by (\ref{R5}%
) i.e. ASEM.

\section{Is ASEM a wormhole?}

The Yilmaz Exponential Metric (YEM) given in (\ref{1.1}) has been shown to
describe a traversable wormhole with a throat located at $r=M$. This was
demonstrated by minimizing the area of the constant-$r$ surface in \cite{E1}%
, where the minimum occurs at $r=M.$ Applying a similar analysis to the
ASEM, the area of a surface at fixed $r$ is given by%
\begin{equation}
A\left( r\right) =\int_{0}^{2\pi }\int_{0}^{\pi }e^{\left( +\frac{2M}{r}%
\right) }e^{\left( -\frac{B^{2}}{2r^{2}}\sin ^{2}\theta \right) }r^{2}\sin
\theta d\theta d\phi ,  \label{2.15}
\end{equation}%
which simplifies to%
\begin{equation}
A\left( r\right) =4\pi r^{2}e^{\left( +\frac{2M}{r}\right) }\int_{0}^{\pi
/2}e^{\left( -\frac{B^{2}}{2r^{2}}\sin ^{2}\theta \right) }\sin \theta
d\theta .  \label{2.16}
\end{equation}%
Using the substitution $\sin ^{2}\theta =u,$ we obtain%
\begin{equation}
A\left( r\right) =4\pi r^{2}e^{\left( +\frac{2M}{r}\right) }\int_{0}^{1}%
\frac{e^{-\frac{B^{2}}{2r^{2}}u}du}{\sqrt{1-u}}.  \label{2.17}
\end{equation}%
This integral can be expressed in terms of the error function, resulting in 
\begin{equation}
A\left( r\right) =-\frac{4i\sqrt{2}\pi ^{3/2}r^{3}e^{\left( \frac{2M}{r}-%
\frac{B^{2}}{2r^{2}}\right) }}{B}\text{erf}\left( \frac{iB}{r\sqrt{2}}%
\right) .  \label{2.18}
\end{equation}%
\begin{figure}[tbp]
\includegraphics[width=100mm,scale=1]{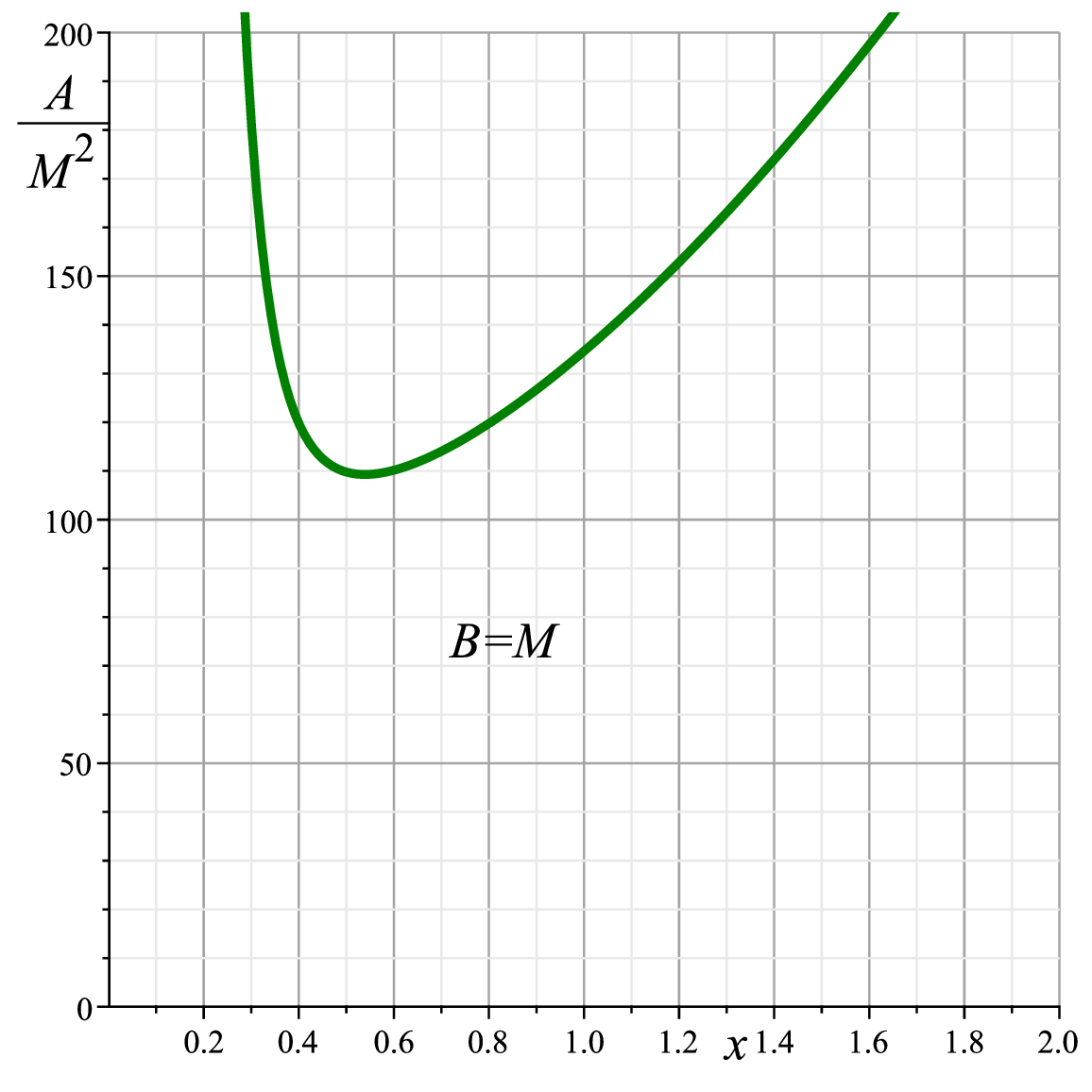}
\caption{The plot of $\frac{A\left( x\right) }{M^{2}}$ for $B=M$ in terms of 
$x=\frac{r}{M}.$ The minimum area is not the actual minimum.}
\label{f1}
\end{figure}
We introduce the dimensionless variable $r=Mx,$ and $B=M$ to plot $\frac{%
A\left( x\right) }{M^{2}}$ as a function of $x$ in Fig. \ref{f1}, which
shows a minimum at $x_{0}=\frac{r_{0}}{M}=0.5389$.

Moreover, the metric functions remain finite and nonzero at $r=r_{0}$.
Therefore, if we define the throat of a wormhole as the location of a
minimum-area surface, the line element in (\ref{2.13}) appears to describe a
wormhole with a throat at $r=r_{0}$. This behavior is reminiscent of the
YEM. However, unlike the spherically symmetric YEM in (\ref{1.1}), the
surface $r=r_{0}$\ in the ASEM does not correspond to a true area minimum in
a general geometric sense. To clarify this, consider the standard procedure
for identifying a minimal-area surface by parametrizing it as $r=R\left(
\theta \right) .$ According to the line element (\ref{2.13}), the
corresponding area functional becomes%
\begin{equation}
A=2\pi \int_{0}^{\pi }\frac{\sqrt{R^{\prime 2}+R^{2}}\sqrt{\xi \left(
R,\theta \right) }}{\psi \left( R\right) }R\sin \theta d\theta .  \label{R7}
\end{equation}%
Changing variables to $u=\cos \theta ,$\ this becomes%
\begin{equation}
A=2\pi \int_{-1}^{1}\frac{\sqrt{\left( 1-u^{2}\right) R^{\prime 2}+R^{2}}%
\sqrt{\xi \left( R,u\right) }}{\psi \left( R\right) }Rdu  \label{R8}
\end{equation}%
where $R^{\prime }=\frac{dR\left( u\right) }{du}$\ and $\xi \left( R\left(
u\right) ,u\right) =e^{\left( -\frac{B^{2}}{R^{2}}\left( 1-u^{2}\right)
\right) }.$\ Varying the functional (\ref{R8}) with respect to $R\left(
u\right) $\ yields the following nonlinear second-order differential
equation 
\begin{multline}
\left( 1-u^{2}\right) R^{4}R^{\prime \prime }+u\left( 1-u^{2}\right) \left(
B^{2}\left( 1-u^{2}\right) -R^{2}\right) R^{\prime 3}+\left( 1-u^{2}\right)
R\left( -B^{2}\left( 1-u^{2}\right) +2MR-3R^{2}\right) R^{\prime 2}-
\label{R9} \\
uR^{2}\left( -B^{2}\left( 1-u^{2}\right) +2R^{2}\right) R^{\prime
}+R^{3}\left( -B^{2}\left( 1-u^{2}\right) +2MR-2R^{2}\right) =0.
\end{multline}%
This equation determines the surface of minimal area for a given $M$ and $B.$%
\ In the case $B=0$, the spacetime reduces to spherically symmetric YEM and (%
\ref{R9}) reduces to%
\begin{equation}
\left( 1-u^{2}\right) R^{4}R^{\prime \prime }-u\left( 1-u^{2}\right)
R^{2}R^{\prime 3}+\left( 1-u^{2}\right) R\left( 2MR-3R^{2}\right) R^{\prime
2}-uR^{2}\left( 2R^{2}\right) R^{\prime }+R^{3}\left( 2MR-2R^{2}\right) =0.
\label{N2}
\end{equation}%
One can see that $R=M$ is an exact solution for (\ref{N2}) confirming that
the YEM is indeed a wormhole geometry with the minimum area located at $r=M.$
In the general configuration when $B\neq 0$, $R=const$\ is not a solution of
(\ref{R9}) implying the 2-parameter ASEM is not representing a wormhole with
the throat located at $r=const.$ Therefore, in contrast to the spherically
symmetric YEM, the surface $r=r_{0}$\ does not define a true wormhole throat
in the case of the ASEM. A similar conclusion was drawn by Bronnikov and
Skvortsova in \cite{Bron}, where they analyzed a particular family of
Zipoy--Voorhees metrics as an alternative to the ASEM. To conclude, we refer
the reader to other known static and axially symmetric wormhole solutions
discussed in \cite{W1,W2,W3,W4}.

\section{Circular Orbits}

In this section, we analyze the circular orbits of both massive and massless
particles, highlighting how the geometric deformation influences the motion
compared to the exponential wormhole.

We begin with the Lagrangian $\mathcal{L}$%
\begin{equation}
2\mathcal{L}=-e^{\left( -\frac{2M}{r}\right) }\left( \frac{dt}{d\sigma }%
\right) ^{2}+e^{\left( +\frac{2M}{r}\right) }\left[ e^{\left( -\frac{B^{2}}{%
r^{2}}\sin ^{2}\theta \right) }\left( \left( \frac{dr}{d\sigma }\right)
^{2}+r^{2}\left( \frac{d\theta }{d\sigma }\right) ^{2}\right) +r^{2}\sin
^{2}\theta \left( \frac{d\phi }{d\sigma }\right) ^{2}\right] .  \label{3.1}
\end{equation}%
From this, we derive the conserved quantities, the energy%
\begin{equation}
E=e^{\left( -\frac{2M}{r}\right) }\left( \frac{dt}{d\sigma }\right) ,
\label{3.2}
\end{equation}%
and the angular momentum%
\begin{equation}
\ell =e^{\left( +\frac{2M}{r}\right) }r^{2}\sin ^{2}\theta \frac{d\phi }{%
d\sigma }.  \label{3.3}
\end{equation}%
Additionally, the equation of motion in the $\theta $-direction is%
\begin{equation}
e^{\left( +\frac{2M}{r}\right) }\cos \theta \left\{ -\frac{B^{2}\sin \theta
e^{\left( -\frac{B^{2}}{r^{2}}\sin ^{2}\theta \right) \left( \left( \frac{dr%
}{d\sigma }\right) ^{2}+r^{2}\left( \frac{d\theta }{d\sigma }\right)
^{2}\right) }}{r^{2}}+\frac{\ell ^{2}e^{\left( -\frac{4M}{r}\right) \left( 
\frac{d\phi }{d\sigma }\right) ^{2}}}{r^{2}\sin ^{3}\theta }\right\} =\frac{d%
}{d\sigma }\left[ e^{\left( \frac{2M}{r}-\frac{B^{2}}{r^{2}}\sin ^{2}\theta
\right) }r^{2}\left( \frac{d\theta }{d\sigma }\right) \right] .  \label{3.4}
\end{equation}%
Unlike in spherically symmetric spacetimes such as the exponential metric, $%
\theta =const.$ is not a trivial solution unless $\cos \theta =0$.
Therefore, the equation is satisfied only in the equatorial plane, i.e., at $%
\theta =\frac{\pi }{2}.$ Using the normalization condition for the
particle's four-velocity $g_{\mu \nu }\frac{dx^{\mu }}{d\sigma }\frac{%
dx^{\nu }}{d\sigma }=\varepsilon =-1,0$ for the massive and massless
particle, respectively, we obtain the main equation of motion 
\begin{equation}
\left( \frac{dr}{d\sigma }\right) ^{2}=e^{\left( +\frac{B^{2}}{r^{2}}\right)
}\left\{ E^{2}-V_{eff}\left( r\right) \right\} ,  \label{3.5}
\end{equation}%
where the effective potential is given by%
\begin{equation}
V_{eff}\left( r\right) =e^{\left( -\frac{2M}{r}\right) }\left( e^{\left( -%
\frac{2M}{r}\right) }\frac{\ell ^{2}}{r^{2}}-\varepsilon \right) .
\label{R10}
\end{equation}%
For a circular orbit at $r=r_{c}$, we impose $\frac{dr}{d\sigma }=0$ which
leads to 
\begin{equation}
E^{2}=V_{eff}\left( r\right) =e^{\left( -\frac{2M}{r_{c}}\right) }\left(
e^{\left( -\frac{2M}{r_{c}}\right) }\frac{\ell ^{2}}{r_{c}^{2}}-\varepsilon
\right) .  \label{3.6}
\end{equation}%
An additional condition for circular orbits is $\frac{d^{2}r}{d\sigma ^{2}}%
=0,$ which implies%
\begin{equation}
\left. \frac{dV_{eff}}{dr}\right\vert _{r=r_{c}}=0.  \label{3.7}
\end{equation}%
Interestingly, this condition is exactly the same as in the exponential
metric considered in \cite{E1}. For massless particles ($\varepsilon =0$),
there is only one circular orbit located at $r_{c}=2M$. However, this orbit
is unstable, as indicated by $V_{eff}^{\prime \prime }\left( r_{c}\right)
<0. $ In contrast, for massive particles ($\varepsilon =-1$), i) Stable
circular orbits exist for $r_{c}\geq \left( 3+\sqrt{5}\right) M$. ii)
Unstable circular orbits exist in the range $2M\leq r_{c}<\left( 3+\sqrt{5}%
\right) M.$ iii) No circular orbits are allowed for $r_{c}<2M$. The energy
and angular momentum of a massive particle in a circular orbit are given by%
\begin{equation}
E^{2}=\frac{r_{c}-M}{r_{c}-2M}e^{\frac{2M}{r_{c}}},  \label{3.8}
\end{equation}%
and%
\begin{equation}
\ell ^{2}=\frac{Mr_{c}^{2}}{r_{c}-2M}e^{\frac{2M}{r_{c}}},  \label{3.9}
\end{equation}%
which clearly show that circular orbits require $r_{c}>2M$.

\section{Connection with the Zipoy-Voorhees (ZV) Metric}

Before concluding this article, we briefly comment on a possible connection
between our solution and the Zipoy--Voorhees (ZV) metric. In addition to the
Schwarzschild metric, which is a one-parameter solution of Einstein's vacuum
equations, there exists a two-parameter generalization known as the
Zipoy-Voorhees (ZV) spacetime, or the $\gamma $-metric \cite{ZV1,ZV2}. Given
that the ZV metric is also static and axially symmetric, it is natural to
ask whether the metric given in (\ref{2.13}) belongs to the ZV class.%
\begin{equation}
ds^{2}=-f\left( r\right) ^{\gamma }dt^{2}+f\left( r\right) ^{\gamma
^{2}-\gamma }g\left( r,\theta \right) ^{1-\gamma ^{2}}\left( \frac{dr^{2}}{%
f\left( r\right) }+r^{2}d\theta ^{2}\right) +f^{1-\gamma }r^{2}\sin
^{2}\theta d\phi ^{2}  \label{C1}
\end{equation}%
where the metric functions are defined by 
\begin{equation}
f\left( r\right) =1-\frac{2m}{r},  \label{C2}
\end{equation}%
\begin{equation}
g\left( r,\theta \right) =1-\frac{2m}{r}+\frac{m^{2}\sin ^{2}\theta }{r^{2}},
\label{C3}
\end{equation}%
and $\gamma >0$\ is a free parameter. The constant $m>0$\ is such that the
ADM mass observed at spatial infinity is given by $M=\gamma m.$\ In the
asymptotic limit $r\rightarrow \infty ,$ the line element reduces to%
\begin{equation}
ds^{2}\rightarrow -\left( 1-\frac{2M}{r}\right) dt^{2}+\frac{dr^{2}}{\left(
1-\frac{2M}{r}\right) }+r^{2}\left( d\theta ^{2}+\sin ^{2}\theta d\phi
^{2}\right) ,\text{ \ as }r\rightarrow \infty .  \label{C4}
\end{equation}%
which is the Schwarzschild metric in standard coordinates. Rewriting the
metric functions in terms of the ADM mass $M$, we obtain 
\begin{equation}
f\left( r\right) =1-\frac{2M}{\gamma r},  \label{C5}
\end{equation}%
and%
\begin{equation}
g\left( r,\theta \right) =1-\frac{2M}{\gamma r}+\frac{M^{2}\sin ^{2}\theta }{%
\gamma ^{2}r^{2}}.  \label{C6}
\end{equation}%
Now, consider the limit $\gamma \rightarrow \infty .$\ In this limit, the ZV
metric (\ref{C1}) reduces to the CCS presented in (\ref{2.13}).

\section{Conclusion}

In this work, we have constructed a family of axially symmetric solutions to
Einstein's field equations that generalize the well-known Yilmaz Exponential
Metric (YEM). Starting from the Curzon--Chazy spacetime as a classical
example of axial symmetry, we reformulated the field equations in spherical
coordinates and derived a two-parameter Axially Symmetric Exponential Metric
(ASEM) that reduces to the YEM in an appropriate limit. This family
interpolates between the non-vacuum YEM and the vacuum Curzon--Chazy (CCS)
geometries. The corresponding energy-momentum tensor encompasses both
physical and exotic matter regimes, depending on the relation between the
parameters $M$ and $B$. Notably, the CCS emerges as the unique, sourceless
solution within this family when $M^{2}=B^{2}$. We also analyzed the
geometric properties of the ASEM and showed that in the special case $B=0$,
the solution meets the criteria for a traversable wormhole, with a throat
located at the finite radius $r_{0}=M$, where the area of constant-$r$
surfaces attains a minimum, as previously reported in \cite{E1}. However,
for nonzero $B$, the angular deformation causes the constant-$r$ surface at $%
r=r_{0}$ to lose its status as a minimal surface in the general geometric
sense, distinguishing it from the spherically symmetric YEM wormhole.
Finding a general solution to Eq. (\ref{R9}) remains an open problem. If
such a solution exists, it would allow for a detailed analysis of the
possible wormhole's throat structure.

Acknowledgement

The author would like to thank Prof. M. Halilsoy, for the useful discussions
on the ZV metric.


\begin{thebibliography}{99}
\bibitem{E1} P. Boonserm, T. Ngampitipan, A. Simpson and M. Visser, Phys.
Rev. D 98, 084048.

\bibitem{P1} A. Papapetrou, Zeitschrift fur Physik, Bd. 139, 518 (1954).

\bibitem{P2} A. Papapetrou, Math. Nachr. 12, 129 (1954).

\bibitem{P3} A. Papapetrou, Math. Nachr. 12, 143 (1954).

\bibitem{Y1} H. Yilmaz, Phys. Rev. 111, 1417 (1958).

\bibitem{Y2} H. Yilmaz, Phys. Rev. Lett. 27, 1399 (1971).

\bibitem{Y3} H. Yilmaz, Ann. Phys. (N.Y.) 81, 179 (1973).

\bibitem{E2} R. E. Clapp, Phys. Rev. D 7, 345 (1973).

\bibitem{E3} N. Rosen, Ann. Phys. (N.Y.) 84, 455 (1974).

\bibitem{E4} P. Rastall, Gravity without geometry, Am. J. Phys. 43, 591
(1975).

\bibitem{E5} A. J. Fennelly and R. Pavelle, University of Western Kentucky
Report No. Print76-0905, 1976.

\bibitem{E6} A. A. Svidzinsky, arXiv: 0904.3155.

\bibitem{E7} C. O. Alley, P. K. Aschan, and H. Yilmaz, arXiv:gr-qc/9506082.

\bibitem{E8} C. W. Misner, Nuovo Cimento B 114, 1079 (1999).

\bibitem{E9} S. L. Robertson, Astrophys. J. 515, 365 (1999).

\bibitem{E10} S. L. Robertson, Astrophys. J. 517, L117 (1999).

\bibitem{E11} Y. Itin, Gen. Relativ. Gravit. 31, 187 (1999).

\bibitem{E12} S. L. Robertson, Astrophys. J. 515, 365 (1999).

\bibitem{E13} K. Watt and C. W. Misner, arXiv:gr-qc/9910032.

\bibitem{E14} V. Zhuravlev and D. Kornilov, Gravitation Cosmol. 5, 325
(1999).

\bibitem{E15} M. Ibison, Classical Quantum Gravity 23, 577 (2006).

\bibitem{E16} M. Ibison, AIP Conf. Proc. 822, 181 (2006).

\bibitem{E17} M. Ibison, Classical Quantum Gravity 23, 577 (2006).

\bibitem{E18} N. Ben-Amots, Found. Phys. 37, 773 (2007).

\bibitem{E19} S. Ayg\"{u}n, \.{I}. Tarhan, and H. Baysal, Astrophys. Space
Sci. 314, 323 (2008).

\bibitem{E20} M. Martinis and N. Perkovic, arXiv:1009.6017.

\bibitem{E21} N. Ben-Amots, J. Phys. Conf. Ser. 330, 012017 (2011).

\bibitem{E22} M. E. Aldama, J. Phys. Conf. Ser. 600, 012050 (2015).

\bibitem{E23} S. L. Robertson, arXiv: 1606.01417.

\bibitem{E24} A. Svidzinsky, Phys. Scr. 92, 125001 (2017).

\bibitem{E25} M. A. Makukov and E. G. Mychelkin, Phys. Rev. D 98, 064050
(2018).

\bibitem{T1} M. S. Morris and K. S. Thorne, Am. J. Phys. 56, 395 (1988).

\bibitem{T2} M. S. Morris, K. S. Thorne, Phys. Rev. Lett. 61, 1446 (1988).

\bibitem{T3} M. Visser, Phys. Rev. D 39, 3182 (1989).

\bibitem{T4} M. Visser, Nucl. Phys. B 328, 203 (1989).

\bibitem{T5} M. Visser, Lorentzian Wormholes: From Einstein to Hawking
(Springer, New York, 1995).

\bibitem{Cur} H. Curzon, Proceedings of the London Mathematical Society 2,
477 (1925).

\bibitem{Cha} J. Chazy, Bull. Soc. Math. France 52, 17 (1924).

\bibitem{Bron} K. A. Bronnikov and M. V. Skvortsova, Grav. Cosmol. 20, 171
(2014).

\bibitem{W1} B. Narzilloev, D. Malafarina, A. Abdujabbarov, B. Ahmedov, and
C. Bambi, Phys. Rev. D 104, 064016 (2021).

\bibitem{W2} G. Cl\'{e}ment, Int. J. Theor. Phys. 23, 335 (1984).

\bibitem{W3} G. Cl\'{e}ment, Gen. Relativ. Gravit. 16, 477 (1984).

\bibitem{W4} G. Cl\'{e}ment, Gen. Relativ. Gravit. 48, 76 (2016).

\bibitem{ZV1} D. M. Zipoy, J. Math. Phys. 7, 1137 (1966).

\bibitem{ZV2} B. H. Voorhees, Phys. Rev. D 2, 2119 (1970).
\end{thebibliography}
\end{document}